# Large Deviation Theory Approach to Fluctuation Theorems and Landauer's Principle through Heat Redefinition

Tatsuaki Tsuruyama[1,2,3]

[1] Department of Physics, Tohoku University, Sendai 980-8578, Japan

[2] Department of Drug Discovery Medicine, Kyoto University, Kyoto 606-8501, Japan

[3] Department of Clinical Laboratory, Kyoto Tachibana University, Kyoto 607-8175, Japan

**Abstract**

Large deviation theory (LDT) provides a mathematical framework to quantify the probabilities of rare events in stochastic systems. In this study, we applied LDT to model a chemical reaction system and demonstrated that the fluctuation theorem for nonequilibrium reaction systems can be derived from the symmetry of the cumulant generating function defined through the rate function. Notably, this derivation does not depend on the assumption of local detailed balance. Furthermore, we redefined heat using this rate function based on information theory and evaluated Landauer's principle, which addresses the minimum energy cost associated with information erasure. These findings show the utility of LDT as a comprehensive framework for analyzing a wide range of nonequilibrium systems.

*Contact author: tsuruyam@ddm.med.kyoto-u.ac.jp

## 1. Text

Large deviation theory (LDT) ie one of the robust mathematical frameworks for quantifying the probabilities of rare events in stochastic physical systems. In particular, it is beneficial for understanding fluctuations far from equilibrium or steady states. LDT applies to a broad range of fields, including statistical mechanics(Touchette, 2009), thermodynamics(Lebowitz, 1999), and dynamical systems(Hu and Wu, 2011). A solid mathematical foundation for the theory was provided in earlier works (Dembo, 2010). One of the foundational results in LDT is Cramér's theorem:

$$P(\chi) \sim e^{-I(X)T} \tag{1}$$

where $P(\chi)$ represents the probability that a physical quantity $\chi$ deviates to a specific value from the steady state. $T$ is a scaling factor, such as the observation time, the system size, or the total concentration. $I(X)$ is the rate function that quantifies how rare a particular deviation is at the steady state.

The role of LDT in physical systems has been extended to more complex systems, including chemical reactions and transport processes(Barato and Seifert, 2015; Bertini et al., 2015; Lebowitz and Spohn, 1999; Rao and Esposito, 2016). In this study, we apply to derive the fluctuation theorem (FT) based on LDT. Using this framework, it becomes possible to advance the information thermodynamic discussion without the assumption of local detailed balance (LDB), which is generally applied in nonequilibrium thermodynamics.

## 2. Results

### 2.1. Chemical Model System

We consider a system where two chemical species react simultaneously in contact with a chemical bath to transmit the information. Let $X$ and $Y$ represent the concentrations of the two chemical species. These species undergo reactions according to the following reaction mechanism:

$$X \leftrightarrow Y \tag{2}$$

In the above formula, $X$ is an inactive chemical species that does not carry information, and $Y$ is an active chemical species that carries information. And we set:

$$X + Y = A \tag{3}$$

*Contact author: tsuruyam@ddm.med.kyoto-u.ac.jp

where $A$ is kept constant. In this model, the forward transition is more frequent in Eq. (2), while the reverse is rare or slow, indicating that there is a directionality with respect to information transmission.. The model reaction system is assumed to be in contact with a chemostat, which continuously supplies substances to maintain a steady-state environment, allowing conversion reactions between $X$ and $Y$ to occur. Such reaction models are commonly accepted in intracellular biochemical reactions(Tsuruyama, 2019). The master equation governing these dynamics is given by:

$$\frac{dP(X,t)}{dt} = k_f(X \to Y)P(X,t) - k_b(Y \to X)P(Y,t) \tag{4}$$

Here, $P(X, t)$ is the probability function of $X$ molecules at time $t$, and $k_f$ (X→Y) and $k_b$ (Y→X) denote the forward and backward reaction rate constants, respectively.

## 2.2. Analyzing the Hamiltonian with Large Deviation Theory

We then aimed to describe the dynamics of the system in terms of a Hamiltonian formalism that describes the energy involved in the transition between X and Y. To achieve this, we introduced $\psi$, which functions as a free energy variable. In particular, it quantifies how likely rare transitions (large deviations) occur compared to the steady-state behavior. In this context: $e^{\psi} -1$ acts as a creation operator, which accounts for an increase in the number of molecules $X$ during a reaction, $e^{-\psi}-1$ acts as an annihilation operator, which represents the reduction in the number of molecules $X$. These operators describe the probabilistic transitions between $X$ and $Y$. Here we set $p(\tau)=X(\tau)/A$ where $\tau$ denotes the reaction time. The chemical master equation (4) can be transformed using a Hamiltonian description:

$$H(p(\tau);\psi) = k_f p(\tau)(e^{\psi} - 1) + k_b p(\tau)(e^{-\psi} - 1) \tag{5}$$

To analyze this Hamiltonian within the framework of LDT, the rate function is defined as follows:

$$I(p(\tau)) = \int_0^{\tau} \left[\psi(p(\tau))\frac{dp(\tau)}{dt} - H(p(\tau);\psi)\right] dt \tag{6}$$

The condition where the Hamiltonian $H(p(\tau), \psi)$ equals zero corresponds to a steady state:

$$H(p(\tau), \psi) = k_f p(\tau)(e^{\psi} - 1) + k_b (1 - p(\tau))(e^{-\psi} - 1) = 0 \tag{7}$$

Solving this with respect to $e^{\psi}$, we obtain the following

*Contact author: tsuruyam@ddm.med.kyoto-u.ac.jp

$$\psi = -\log \frac{k_f p(\tau)}{k_b(1-p(\tau))} \tag{8}$$

representing the ratio of the forward and reverse reactions in a chemical process under nonequilibrium conditions. By setting the Hamiltonian $H(p(\tau), \psi)$ zero in Eq. (6), the rate function is simply given by:

$$I(p(\tau)) = \int_0^\tau \left[\psi \frac{dp(\tau)}{dt}\right] dt \tag{9}$$

From Eqs (8) and (9), we obtain the rate function as:

$$I(p(\tau)) = -\int_0^\tau \left[\log \frac{k_f p(\tau)}{k_b(1-p(\tau))} \frac{dp(\tau)}{dt}\right] dt = -\int_0^\tau \left[\log \frac{k_f X(\tau)}{k_b Y(\tau)} \frac{dp(\tau)}{dt}\right] \tag{10}$$

Thus, the rate function incorporates the product of the chemical affinity and the reaction rate as the integrand, which is equivalent to entropy production, $\sigma$. Changing the integral variable from $t$ to $p(\tau)$, we obtain:

$$I(p(\tau)) = -\int_0^\tau \left[\log \frac{k_f p(\tau)}{k_b(1-p(\tau))}\right] dp(\tau). \tag{11}$$

Furthermore, rewriting Eq. (10) gives:

$$I(p(\tau)) = \int_0^\tau \left(k_f p(\tau) - k_b(1-p(\tau))\right) \log \frac{k_f p(\tau)}{k_b(1-p(\tau))} dt = \sigma \tag{12}$$

At $\tau = 0$, the chemical reaction system is in equilibrium. Subsequently, the reaction progresses due to the supply of substances from the bath. For example, we consider the system near equilibrium. Let the equilibrium values of $p(\tau)$ be $p^*(0)$ at $t=0$ using the asterisk symbol representing the equilibrium state, and set the change $\Delta p(\tau)$:

$$p(\tau) = p(\tau)^* + \Delta p(\tau) \tag{13}$$

In this case,

$$\frac{k_f p(0)^*}{k_b(1-p(0)^*)} = 1 \tag{14}$$

and we have:

*Contact author: tsuruyam@ddm.med.kyoto-u.ac.jp

$$\frac{k_f p(\tau)}{k_b(1-p(\tau))} = \frac{p(0)^*}{1-p(0)^*} \frac{Y}{X} \tag{15}$$

Substituting this into Eq. (11), we obtain:

$$I(X) = p(\tau) \log \frac{p(\tau)}{p(0)^*} + (1-p(\tau)) \log \frac{1-p(\tau)}{1-p(0)^*} \tag{16}$$

The right-hand side of Eq. (16) has the negative form of the Kullback-Leibler divergence(KLD), $D_{KL}$ (Kullback and Leibler, 1951). Here we aimed to obtain the cumulant generating function (CGF) $\Phi$ with an arbitary parameter α. CGF can be defined using a control parameter $\lambda$ :

$$\Phi(\lambda, p(\tau)) = \sup_{p(\tau)} \left( \lambda p(\tau) - k_f \tau \left[ \left( p(\tau) \log \frac{p(\tau)}{p(0)^*} + (1-p(\tau)) \log \frac{1-p(\tau)}{1-p(0)^*} \right) \right] \right) \tag{17}$$

Likewise, for the reverse reaction, the CGF can be defined using a control parameter $\lambda\dagger$ where the dagger symbol represents the reverse orientation:

$$\Phi(\lambda^\dagger, 1-p(\tau)) = \sup_{p(\tau)} \left( \lambda^\dagger p(\tau) - k_b \tau \left[ \left( p(\tau) \log \frac{p(\tau)}{p(0)^*} + (1-p(\tau)) \log \frac{1-p(\tau)}{1-p(0)^*} \right) \right] \right) \tag{18}$$

In this case, $k_f \tau = X$ and $k_b \tau = Y$, which are scale factors. Differentiating both CGFs with respect to the control parameters and setting them to zero, we have:

$$\lambda = k_f \tau \left[ \left( \log \frac{p(\tau)}{p(0)^*} - \log \frac{1-p(\tau)}{1-p(0)^*} \right) \right], \lambda^\dagger = k_f \tau \left[ \left( \log \frac{1-p(\tau)}{1-p(0)^*} - \log \frac{p(\tau)}{p(0)^*} \right) \right]$$

CGFs are subsequently given by input above result to Eqs. (17) and (18):

$$\Phi(\lambda^*, p(\tau)) = -k_f \tau \left[ (1-p(\tau)) \log \frac{1-p(\tau)}{1-p(0)^*} \right], \Phi(\lambda^{*\dagger}, 1-p(\tau)) = -k_b \tau \left[ \left( p(\tau) \log \frac{p(\tau)}{p(0)^*} \right) \right] \tag{20}$$

These can be approximated using Eq.(13) as $\Phi(\lambda^*, p(\tau)) = -k_f \Delta p/(1-p(0)^*)$ and $\Phi(\lambda^\dagger, 1-p(\tau)) = -k_b \Delta p/p(0)^*$. Therefore, $\Phi(\lambda^*, p(\tau))/\Phi(\lambda^\dagger, 1-p(\tau)) = k_f p(0)^*/k_b (1-p(0)^*) = 1$. This result definitively represents the symmetry of CGF with respect to the exchange of $p(\tau)$ with $1-p(\tau)$. Moreover, by substituting $p(\tau) = \frac{X}{A}, 1-p(\tau) = \frac{A-X}{A}$ into $\Phi(\lambda^*, p(\tau)) = \Phi(\lambda^{*\dagger}, 1-p(\tau))$, we can rewrite it as:

*Contact author: tsuruyam@ddm.med.kyoto-u.ac.jp

$$\Phi(\lambda^*, X) = \Phi(\lambda^{*\dagger}, A - X)$$

(21)

This expression highlights the symmetry of the CGF. Next, we consider the path taken by the reaction system from time $t = 0$ to $\tau$, which we denote $x_\tau := \{x(t)\}_{t=0}^\tau$. This vector form represents the system status. Let $\lambda$ be the control parameter, corresponding to the progression of the reaction, and its time dependence is given by $\lambda_\tau := \{\lambda(t)\}_{t=0}^\tau$. The probability of the system following the path $x_\tau$ under the control parameter $\lambda_\tau$ is denoted as $p(x_\tau|\lambda_\tau)$. Using these notations, we can define a quasi-static process where the system's initial state is $x(0)$, and the probability density for the realization of the path $x_\tau$ is given by $p(x_\tau|x(0), \lambda_\tau)$. In addition, we consider time reversal path when the time-reversed control parameter $\lambda^\dagger_\tau$ is applied. The initial condition is $x^\dagger(0) := x(\tau)$, the probability density for realizing the time-reversed path $x_\tau^\dagger$ is $p(x^\dagger_\tau|x^\dagger(0), \lambda_\tau^\dagger)$. In particular, the forward path $x_\tau$ is the path where $Y$ is produced, while the reverse path $x^\dagger_\tau$ is the path where $X$ is produced. Then let consider the CGF using the exponential of the rate function, $exp(-I(x_\tau|\lambda_\tau))$ as follows:

$$\Phi(X) := -\log \int_{x(0)}^{x(\tau)} Dx_\tau p(x_\tau|\lambda_\tau) \exp\{-XI(x_\tau|\lambda_\tau)\}$$

(22)

For the reverse path integral, let $x^\dagger_\tau := \{x(t-\tau)\}_\tau^{t=0}$. The path integral element satisfies $Dx_\tau = Dx^\dagger_\tau$. The probability factor in the path integral is denoted as $p^\dagger(x_\tau|\lambda_\tau)$. From Eq. (12), the rate function in the model satisfies:

$$I(x_\tau|\lambda_\tau) = -I^\dagger(x^\dagger_\tau|\lambda^\dagger_\tau) = \sigma$$

(23)

Using Eq. (23), the CGF for the reverse path can be expressed with the total concentration $A$ and $X$ (APPENDIX)(Kawasaki, 2003):

$$\Phi^+(A-X) = -\sum_i \log \int_{x(0)}^{x(t)} Dx_\tau p(x^\dagger_\tau|\lambda^\dagger_\tau) \exp\{-(A-X)I^\dagger(x^\dagger_\tau|\lambda^\dagger_\tau)\}$$

$$= -\sum_i \log \int_{x(0)}^{x(t)} Dx_\tau p(x^\dagger_\tau|\lambda^\dagger_\tau) \exp\{AI(x_\tau|\lambda_\tau)\}\exp\{-XI(x_\tau|\lambda_\tau)\}$$

(24)

From Eq. (21), Eq. (22), and Eq. (24) we obtain the following expression (Crooks, 1999):

*Contact author: tsuruyam@ddm.med.kyoto-u.ac.jp

$$p(x_\tau|\lambda_\tau) = p(x^\dagger{}_\tau|\lambda^\dagger{}_\tau)exp\{AI(x_\tau|\lambda_\tau)\} \qquad (25)$$

Eq. (25) expresses the detailed FT. Thus, we successfully derived FT using LDT without assuming LDB.

## 2. 3. Consideration of the Definition of Heat and Landauer's Principle

Based on the result discussed above, we aimed to redefine the generated heat $q$ during the chemical reaction described by Eq. (2), using the entropy production as follows(Prigogine and George, 1983):

$$\sigma = \Delta S - \beta q \qquad (26)$$

where $\Delta S$ denotes the change of the Shannon entropy of the system. Alternatively, the heat dissipated to the surroundings, $q_{ex} = -q$, can be expressed as follows:

$$\beta q_{ex} = \sigma - \Delta S \qquad (27)$$

Rearranging Eq. (26), we have:

$$\beta q = -D_{KL} + (S(p) - S(p^*)) \qquad (28)$$

where

$$S(p) = -p \log p - (1-p) \log(1-p) \qquad (29)$$

and

$$S(p^*) = -p^* \log p^* - (1-p^*) \log(1-p^*) \qquad (30)$$

In Eq. (28), we used $\sigma = D_{KL}$ in Eq. (12). The time parameter $\tau$ of $p$ has been omitted in above. We then consider a binary code consisting of $X$ and $Y$ of the model system. For instance, when the probability distribution is given by $P^*(X, Y) = \{1/2, 1/2\}$, the Shannon entropy of this state, $S(p^*)$, is given by log 2. When using Eq. (30), $P^* = (k_b/(k_f+k_b), k_f/(k_f+k_b)) = (p^*, 1-p^*)$, the entropy is given by $-p^*$ log $p^*$- $(1-p^*)$ log $(1-p^*)$. Here, let us consider the heat dissipated during the erasure of memory, specifically in relation to Landauer's principle. This principle suggests that the minimum heat dissipation required to erase 1 bit in a symmetric potential and provides the critical link between information theory and nonequilibrium thermodynamics [14, 15]. After the erasure, the final state is a deterministic distribution, $P = \{1,0\}$, or $\{0, 1\}$ and the entropy of this state, $S(p)$, is 0. In this case, the Shannon entropy change is:

$$\Delta S = S(p(\tau)) - S(p(\tau)^*) = -\log 2 \qquad (31)$$

*Contact author: tsuruyam@ddm.med.kyoto-u.ac.jp

In general form, $\Delta S = -p^* \log p^* - (1-p^*) \log (1-p^*)$. Accordingly, the heat dissipation in Eq. (28) can be expressed as:

$$\beta q = -\log 2 - D_{KL} \tag{32}$$

Therefore, heat dissipation can be immediately demonstrated from Eq. (32) as follows:

$$\beta q_{ex} = \log 2 + D_{KL} \geq \log 2 \tag{33}$$

We used $D_{KL} \geq 0$. In the general form, $q = -p^* \log p^* - (1-p^*) \log (1-p^*) + D_{KL}$. In the equilibrium state where $p(t) = p(t)^*$, the entropy production becomes zero ($\sigma = D_{KL}(p||p*) = 0$) and Eq. (33) becomes $\beta q_{ex} = \log 2$, the well-known expression of Landauer's principle. Therefore, it can be understood that Eq. (33) represent a generalization form of Landauer's principle to the heat dissipated during memory erasure from nonequilibrium states. By substituting $\sigma$ in Eq. (16) into the right-hand side of Eq. (27) and incorporating S(p*) defined in Eq. (30), we can obtain the general expression for the dissipated heat: :

$$\beta \Delta q_{ex} = -p \log p^* - (1-p) \log(1-p^*) := H(p, p^*) \tag{34}$$

Notably, the right-hand side of Eq. (34) takes the form of cross-entropy..

## 3. Discussion

In this study, we reported a novel method for deriving FT without the assumption of LDB, based on LDT. FT is theoretically based on the Hamiltonian system under canonical distribution(Sagawa and Ueda, 2012). This is why we adopted a single-molecule reaction model with a Hamiltonian. This theoretical framework allowed for an evaluation of the heat dissipated during memory erasure from an information science perspective, thereby deriving Landauer's principle in Eq. (33) (Andrieux and Gaspard, 2004; Chong et al., 2010; Lebowitz and Spohn, 1999; Seifert, 2008). Moreover, we demonstrated that the heat dissipation is equivalent to cross-entropy.

The result in Eq. (34) is critical for evaluating the energy efficiency of data reset operations in information devices, such as hard drives and memory chips, and for optimizing performance (M. Esposito, 2011) and provides insights into molecular motor movement associated with these reset operations(Toyabe et al., 2010).

*Contact author: tsuruyam@ddm.med.kyoto-u.ac.jp

The findings highlight LDT as a robust and practical mathematical framework for analyzing nonequilibrium information transmission by chemical reaction. We will aim to investigate whether this framework can be extended to encompass broader nonequilibrium reaction networks and phenomenon, including phase transitions and complex nonlinear interactions in future study.

## 4. Conclusions

In this study, we derived FT for chemical reaction systems using LDT, without assuming LDB. Furthermore, we proposed a theoretical derivation of Landauer's principle along with its general expression given by cross entropy.

**Acknowledgements**

This study was supported by a Grant-in-Aid from the Ministry of Education, Culture, Sports, Science, and Technology of Japan (Synergy of Fluctuation and Structure: Quest for Universal Laws in Nonequilibrium Systems, P2013-201). We thank Kenichi Yoshikawa of Kyoto University and Doshisha University for their valuable advice.

**Appndix**

**Caliculation of CGF**

The calculation process of equation (35) is re-shown below.

$$\Phi(X) := -\log \int_{x(0)}^{x(t)} Dx_\tau (x_\tau^\dagger | \lambda_\tau^\dagger) \exp(-XI(x_\tau|\lambda_\tau)) \tag{35}$$

Similarly, the complementary function $\Phi^+(A - X)$ is defined as

$$\Phi^+(A - X) = -\sum_i \log \int_{x^\dagger(t)}^{x^\dagger(\tau-t)} Dx_\tau^\dagger p(x_\tau^\dagger | \lambda_\tau^\dagger) \exp\left(-(A-X)I^\dagger(x_\tau^\dagger|\lambda_\tau^\dagger)\right)$$

$$= -\sum_i \log \int_{x^\dagger(t)}^{x^\dagger(\tau-t)} Dx_\tau^\dagger p(x_\tau^\dagger | \lambda_\tau^\dagger) \exp\left(-AI^\dagger(x_\tau^\dagger|\lambda_\tau^\dagger)\right) \exp\left(XI^\dagger(x_\tau^\dagger|\lambda_\tau^\dagger)\right)$$

*Contact author: tsuruyam@ddm.med.kyoto-u.ac.jp

$$= -\sum_i log \int_{x(0)}^{x(t)} D\boldsymbol{x}_\tau p(\boldsymbol{x}_\tau{}^\dagger|\boldsymbol{\lambda}_\tau{}^\dagger) exp\left(AI(\boldsymbol{x}_\tau|\boldsymbol{\lambda}_\tau)\right) exp\left(-XI(\boldsymbol{x}_\tau|\boldsymbol{\lambda}_\tau)\right)$$

(36)

Furthermore, from Eq. (35), (36), the following relationship holds:

$$p(\boldsymbol{x}_\tau|\boldsymbol{\lambda}_\tau) = p(\boldsymbol{x}_\tau{}^\dagger|\boldsymbol{\lambda}_\tau{}^\dagger) exp\left(AI(\boldsymbol{x}_\tau|\boldsymbol{\lambda}_\tau)\right)$$

(37)

Thus, we derive Equation (37), which corresponds to the FT.

**Refernces**

*Contact author: tsuruyam@ddm.med.kyoto-u.ac.jp

*Contact author: tsuruyam@ddm.med.kyoto-u.ac.jp